\documentclass[11pt]{article}

\usepackage{scicite}
\usepackage{graphicx}
\usepackage{lineno}
\usepackage{amsmath}
\usepackage{url}
\usepackage[labelfont=bf]{caption}

\usepackage{times}

\newenvironment{sciabstract}{%
\begin{quote} \bf}
{\end{quote}}

\title{New Physics-Based Early Warning Signal shows AMOC is on Tipping Course \\ \vspace{1cm}
\small{\emph{First simulated AMOC tipping event in a state-of-the-art climate model}}} 

\author
{Ren\'e M. van Westen$^{1\ast}$, Michael Kliphuis$^{1}$ and Henk A. Dijkstra$^{1}$\\
\\
\normalsize{$^{1}$Institute for Marine and Atmospheric research Utrecht, Utrecht University,}\\
\normalsize{Princetonplein 5, 3584 CC Utrecht, the Netherlands}\\
\\
\normalsize{$^\ast$To whom correspondence should be addressed; E-mail: r.m.vanwesten@uu.nl}
}

\date{}

\begin{document} 


\baselineskip20pt


\maketitle


\begin{sciabstract}
One of the most prominent climate tipping elements is the Atlantic Meridional 
Overturning Circulation (AMOC), which can potentially collapse due to the input of fresh 
water in the North Atlantic. Although AMOC collapses have been induced in state-of-the-art 
Global Climate Models (GCMs) by strong freshwater forcing, an AMOC tipping event has 
so far not been found.  Here, we show results of the first AMOC tipping  event  in 
such a GCM, the Community Earth System Model, including  the large climate impacts 
of the associated AMOC collapse. 
Using these results, we develop a new  physics-based and observable  early warning  signal 
of AMOC tipping:  the minimum  of  the  AMOC  induced  freshwater transport at the southern  
boundary of the Atlantic.   Available observational and reanalysis data indicate that the present-day 
AMOC is  on route to tipping. The new early warning signal is a useful alternative to  classical 
statistical ones which, when applied to our simulated AMOC tipping event, turn out to be 
sensitive to the analysed time interval before tipping. 
\end{sciabstract}


\section*{Introduction}

The Atlantic Meridional Overturning Circulation (AMOC) effectively transports heat and salt 
through the global ocean \cite{Johns2011} and strongly modulates regional and  global
climate.   Continuous  section measurements  of the AMOC, available  since 2004 
at  26$^{\circ}$N  from the RAPID-MOCHA array   \cite{Srokosz2015}, have shown that 
the AMOC strength has decreased by a  few Sverdrups (1 Sv = $10^6$ m$^3$s$^{-1}$) from  
2004 to 2012 and thereafter it has  strengthened  \cite{Moat2020} again.   Longer time 
scale variability  of the AMOC strength,  estimated   by using sea surface temperature 
(SST) time series based on `fingerprint'  patterns  \cite{Ceasar2018}, indicates that the 
AMOC weakened   by $3 \pm 1$~Sv since about 1950. From  proxy records, it has been 
suggested that the AMOC is currently in its  weakest state in over a millennium 
\cite{Ceasar2021}. 

The AMOC has been labelled as  one of the tipping elements in the climate system
\cite{Lenton2008,Armstrong2022}, indicating that it may undergo a relatively rapid change 
under a slowly developing forcing.   The AMOC is particularly sensitive to the ocean's  freshwater 
forcing, either through the surface freshwater flux (e.g., precipitation) or by input of fresh 
water  due to river run-off or ice melt (e.g., from the Greenland Ice Sheet).
Although no AMOC tipping  has been found in  historical observations, there is much 
evidence from proxy records that abrupt AMOC changes have occurred in the geological 
past during the so-called Dansgaard-Oeschger events \cite{Rahmstorf2002, Henry2016, 
Lynch-Stieglitz2017}. 

Classical early warning indicators, such as the increase in the variance and/or the 
(lag-1) auto-correlation, when   applied to SST-based  
time series,  suggest that the present-day AMOC approaches a tipping point before the end of this century
\cite{Boers2021, Ditlevsen2023}.   Apart from the fact that the SST-based AMOC fingerprints may 
not represent the  AMOC behaviour adequately, many (statistical) assumptions are required to
estimate the approaching AMOC tipping point \cite{Ditlevsen2010, Kuehn2011, Qin2018, Ditlevsen2023}. 
Hence, there is strong need for a more 
physics-based, observable, and reliable early warning indicator which characterises the  AMOC tipping point.

\section*{Results}
\section*{AMOC Collapse}

\begin{figure}[h!]

\hspace{-1cm}
\includegraphics{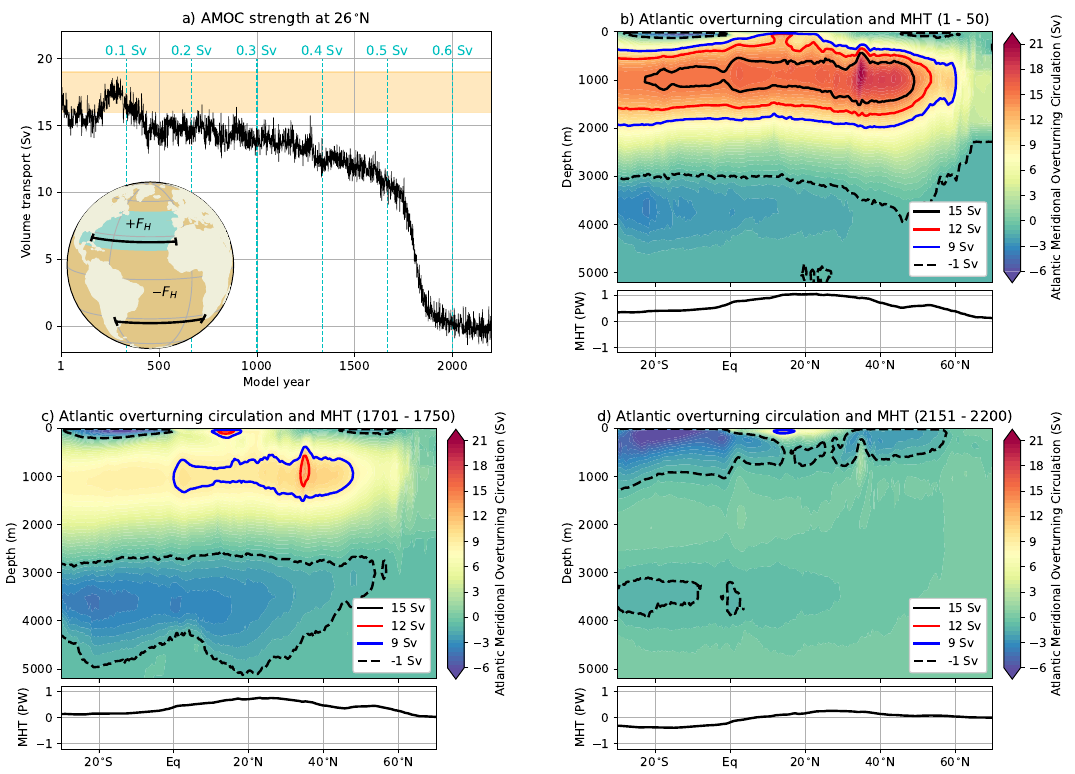}

\caption{\textbf{AMOC collapse.}
(a): The AMOC strength at 1,000~m and 26$^{\circ}$N,
where the yellow shading indicates observed ranges \cite{Smeed2018, Worthington2021}.
The cyan-coloured lines indicate the magnitude of $F_H$.
Inset: The hosing experiment where fresh water is added to the ocean surface between 20$^{\circ}$N -- 50$^{\circ}$N in the Atlantic Ocean ($+F_H$) 
and is compensated over the remaining ocean surface ($-F_H$).
The black sections indicate the 26$^{\circ}$N and 34$^{\circ}$S latitudes over which the AMOC strength and $F_{\mathrm{ovS}}$ are determined, respectively.
(b -- d): The AMOC streamfunction ($\Psi$) and the Atlantic meridional heat transport (MHT, see also Figure~\ref{fig:Figure_S1}) 
for model years 1 -- 50, 1,701 -- 1,750 and 2,151 -- 2,200. The contours indicate the isolines of $\Psi$ for different values.}
\label{fig:Figure_1}
\end{figure}

To develop such an  early warning indicator,  we performed a  targeted simulation to find 
an AMOC tipping event  in the Community Earth System Model (CESM). Specifically, we use
the f19\_g16 configuration of CESM1.0.5, with  horizontal resolutions of  1$^{\circ}$ 
for the  ocean/sea ice  and 2$^{\circ}$ for the atmosphere/land components, which has been 
used in the Coupled Model Intercomparison Project (CMIP), phase 5. We start from a statistical 
equilibrium solution of a pre-industrial control simulation and keep greenhouse gas, solar and 
aerosol forcings constant to pre-industrial levels during the simulation.  A quasi-equilibrium 
approach  \cite{Rahmstorf2005, Hu2012} is followed, by adding a slowly-varying freshwater flux 
anomaly $F_H$ in the North Atlantic over  the region between latitudes  20$^{\circ}$N and 
50$^{\circ}$N. This  freshwater flux anomaly is  compensated  over the rest of the domain  as 
shown in the inset of  Figure~\ref{fig:Figure_1}a.  We linearly increased the freshwater flux 
forcing with a rate of  $3 \times 10^{-4}$~Sv~ yr$^{-1}$ until model year~2,200 where a 
maximum of $F_H = 0.66$~Sv is reached. This is a computationally highly  expensive 
calculation and so cannot easily be repeated for a suite of different GCMs. 

Under increasing freshwater forcing, we find a gradual decrease (Figure~\ref{fig:Figure_1}a) 
in the AMOC strength (see Methods).  Natural variability dominates the AMOC strength in the 
first 400~years but, after model year~800,  a clear negative trend appears due to the increasing freshwater 
forcing.  Then, after 1,750~years of model integration, we find an abrupt AMOC collapse.
The AMOC strength is about 10~Sv in model year~1,750  and decreases to 2~Sv 100~years later 
(model year~1,850) and eventually  becomes slightly  negative after model year~2,000. 
Such a transient AMOC response (model years 1,750 -- 1,850) is spectacular considering 
the slow change in the freshwater forcing (i.e., $\Delta F_H = 0.03$~Sv).  The characteristic  
meridional overturning circulation and associated northward heat transport in   the Atlantic  
Ocean have decreased to nearly zero and by 75\% (at 26$^\circ$N), respectively,  
after model year~2,000 (Figures~\ref{fig:Figure_1}b,c,d and \ref{fig:Figure_S1}a). 

This result differs substantially from earlier model simulations with GCMs which have employed 
extremely large freshwater forcing (e.g., 1~Sv per year over 50$^{\circ}$N -- 70$^{\circ}$N \cite{Orihuela2022}) or large initial 
salinity perturbations \cite{Mecking2017}. The AMOC in these simulations is a direct response 
to the very strong forcing, whereas in our model simulations, which are more akin the simulations 
in Earth System Models of Intermediate Complexity  \cite{Rahmstorf2005}, the collapse is purely 
a response due to internal feedbacks. This can be quantified by looking at the AMOC change 
per cumulative change in North Atlantic freshwater forcing which is about 
$\frac{8~\mathrm{Sv}}{1.5~\mathrm{Sv}} = 5.3$ in our hosing simulations 
(model years~1,750 -- 1,850) and about $\frac{18~\mathrm{Sv}}{50~\mathrm{Sv}} = 0.36$ 
for the strong forcing of 1~Sv per year over a 50-year period for the results in 
\cite{Orihuela2022}. Also based on the change in the AMOC per forcing change 
(here about 8~Sv due to a change in 0.03~Sv surface freshwater flux), it is 
clear that we found an AMOC tipping event \cite{Lenton2008} in the CESM 
simulation which is the first one found in a state-of-the-art GCM. 

The differences in important ocean observables  between the two different AMOC states  
(averages over model years~2,151 -- 2,200 minus years~1 -- 50) are presented 
in Figure~\ref{fig:Figure_S2}.  Figure~\ref{fig:Figure_S2}a shows a cooling of the Northern
Hemispheric SSTs when the 
AMOC collapses, with SST differences as large as 10$^{\circ}$C near Western Europe.
On the contrary, the SSTs in the Southern Hemisphere increase due to the collapse  resulting 
in  a  distinct see-saw pattern between the hemispheres \cite{Stocker1998}.  This pattern arises 
from the reduced  meridional heat exchange between the hemispheres (Figure~\ref{fig:Figure_S1}).
The North Atlantic upper 100~m salinities are also strongly influenced under the  AMOC collapse 
(Figure~\ref{fig:Figure_S2}b). Note that  salinities outside of the Atlantic have increased 
partly due to the freshwater flux compensation used in  the set-up of  the quasi-equilibrium  
experiment. From the changes in the annual maximum mixed-layer depth 
(Figure~\ref{fig:Figure_S2}c), it can be deduced that deep convection ceases in the North Atlantic 
(around Greenland) and convection increases in the Southern Ocean,  in accordance with the 
reversed AMOC  state (Figure~\ref{fig:Figure_1}d).  The weakening of the AMOC 
results, via geostrophic balance, in dynamic sea-level rise in the Atlantic Ocean 
(Figure~\ref{fig:Figure_S2}d) and some coastal regions experience more than 70~cm 
of dynamic sea-level rise. 

\subsection*{Climate Impacts}

The SST changes due to AMOC collapse also affect the atmosphere  and global sea-ice 
distribution. The atmospheric responses (Figure~\ref{fig:Figure_S3}) consist of a see-saw 
pattern in the 2-meter surface temperature, a southward  Inter-Tropical Convergence Zone 
(ITCZ) shift, and the 
strengthening of the Hadley Cell in the Northern Hemisphere.   The stronger meridional 
temperature gradient over the Northern Hemisphere  amplifies the subtropical jet,  while 
the opposite happens in the Southern Hemisphere. During the gradual AMOC weakening 
over the first 1,400~model years,  there are no significant trends  ($p > 0.05$, two-sided t-test \cite{Santer2000}) in the
global mean surface temperature nor in the global sea-ice area. 
Under the AMOC collapse,  the Arctic (March) sea-ice pack extends down to 50$^{\circ}$N and there is a gradual retreat of the 
Antarctic (September) sea-ice pack (Figure~\ref{fig:Figure_S4}). The vast expansion of the 
Northern Hemispheric sea-ice pack amplifies further Northern Hemispheric cooling via the 
ice-albedo feedback.    These findings are qualitatively similar to those in  \cite{Orihuela2022} 
in which AMOC is strongly  weakened to 3 -- 4~Sv.

\begin{figure}[h!]

\begin{tabular}{c}

\hspace{-2.5cm}
\includegraphics[width=1.3\columnwidth, trim = {2.5cm 1cm 0cm 2cm}, clip]{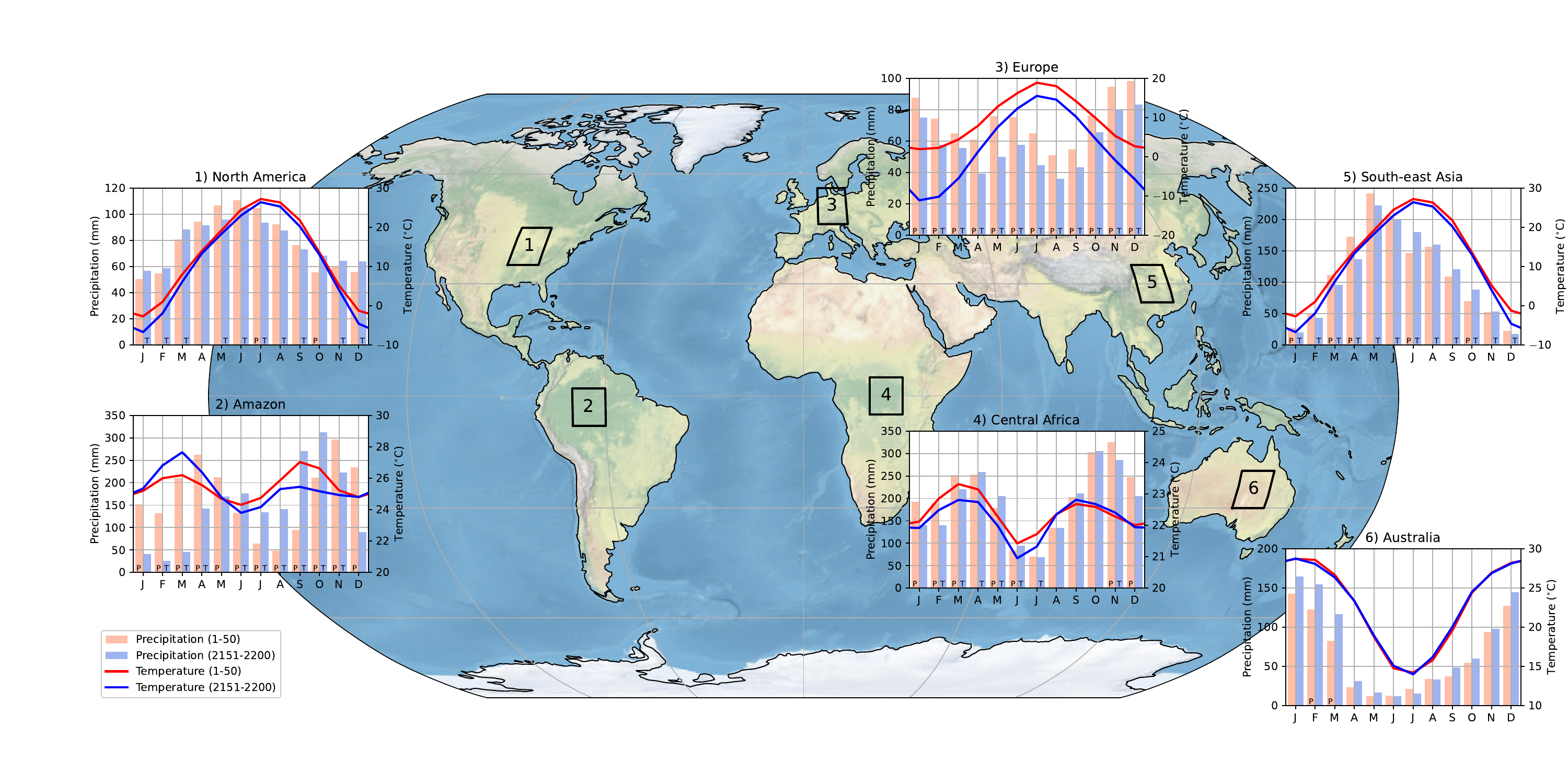}
\end{tabular}
\caption{\textbf{Climograph for different regions.}
The climograph for 6~different regions (spatial average over the $10^{\circ} \times 10^{\circ}$ boxes), where the bars indicate the monthly precipitation and the curves the monthly temperatures.
The climograph is determined over model years~1 -- 50 (red bars and curves) and model years~2,151 -- 2,200 (blue bars and curves). Note the different vertical ranges for each climograph.
The letters~P and T in the bars indicate significant ($p < 0.05$, two-sided Welch's t-test) monthly differences for precipitation and temperature, respectively.}

\label{fig:Figure_2}
\end{figure}

The aforementioned ocean, atmosphere and sea-ice responses strongly influence the regional 
climates across the globe (Figure~\ref{fig:Figure_2}). The European climate is significantly 
different after the AMOC collapse, whereas for other regions only specific months undergo 
significant changes. The Amazon Rainforest also shows a drastic change in their precipitation 
patterns due to ITCZ shifts and the dry season becomes the wet season and vice versa.
These AMOC-induced precipitation changes could severely disrupt the ecosystem of the Amazon 
Rainforest \cite{Hirota2011, Boers2017,  Armstrong2022} and potentially lead to cascading  
tipping  \cite{Dekker2018, Wunderling2021, Klose2021}. The Northern Hemisphere 
shows cooler temperatures after the AMOC collapse while the opposite is true for the 
Southern Hemisphere,  although not all changes are significantly different (due to large 
inter-annual variability). 

\begin{figure}[h!]
\hspace{-2cm}
\includegraphics{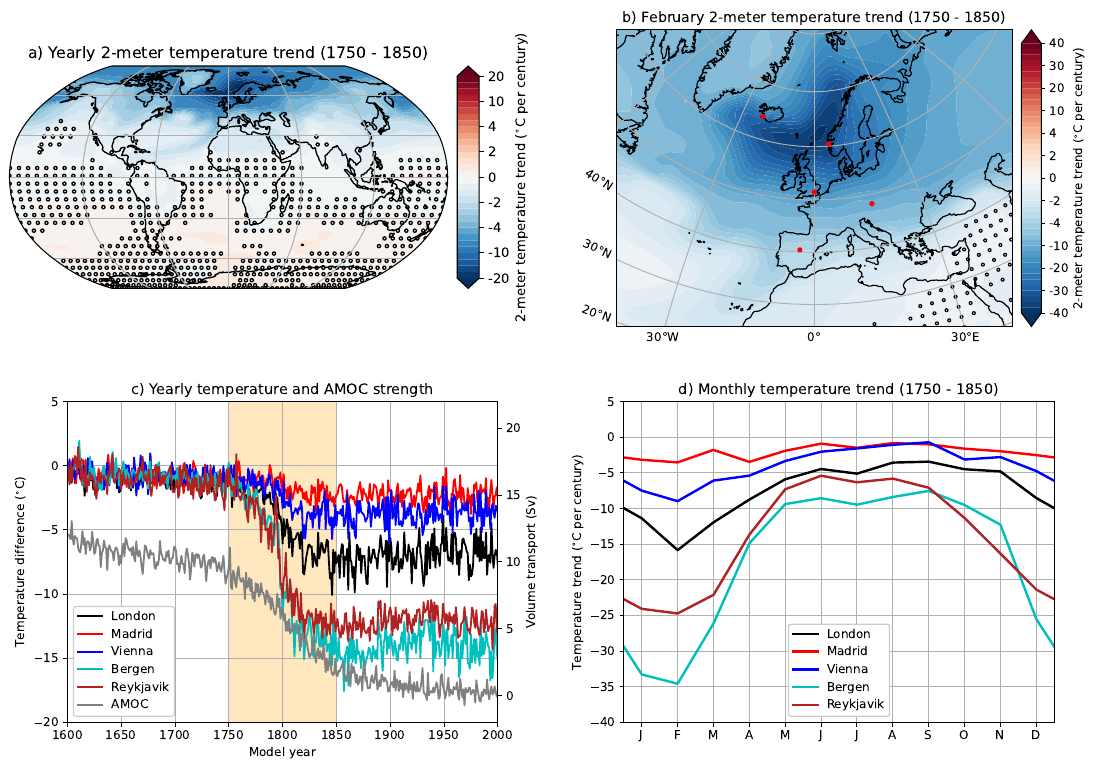}
\caption{\textbf{Surface temperature response during AMOC collapse.}
(a): The yearly-averaged 2-meter surface temperature trend (model years~1,750 -- 1,850), 
the markers indicate non-significant trends ($p \geq 0.05$, two-sided t-test \cite{Santer2000}). 
(b): Similar to panel~a, but now for the February 2-meter surface temperature trend. 
The red dots indicate 5~different cities used in panels~c and d.
Note the different colourbar ranges between panels~a and b.
(c): The temperature difference (w.r.t. model year~1,600) for 5~different cities, including the AMOC strength.
The trends are determined over model~years 1,750 -- 1,850 (yellow shading) during which the AMOC strength strongly decreases.
(d): The monthly temperature trends for the 5~different cities.}

\label{fig:Figure_3}
\end{figure}

The European climate is greatly  affected   (Figure~\ref{fig:Figure_3}a) under the AMOC 
collapse. Note that the corresponding changes occur  within a relatively short period (model 
years~1,750 -- 1,850) and under a very small change in 
surface freshwater forcing.  The yearly-averaged atmospheric surface temperature trend  
exceeds 1$^\circ$C per decade over  a broad region in northwestern Europe  and for 
several European cities, temperatures are found to drop by 5$^\circ$C to 15 $^\circ$C
(Figure~\ref{fig:Figure_3}c).   The  trends are even  more striking when considering 
particular  months (Figure~\ref{fig:Figure_3}b). As an example, February temperatures 
for Bergen (Norway)  will drop by about  3.5$^\circ$C per decade (Figure~\ref{fig:Figure_3}d). 
These relatively strong temperature trends  are associated with the sea-ice albedo feedback 
through the vast expansion of the Arctic sea-ice  pack. 

\newpage

\subsection*{Physics-based Early Warning Indicator}

From idealised ocean-climate models it has been suggested  that the freshwater 
transport of the AMOC  at 34$^{\circ}$S,  indicated by   $F_{\mathrm{ovS}}$ (see Methods), 
is  an important indicator  of AMOC stability \cite{Rahmstorf1996, DeVries2005, 
Dijkstra2007, Huisman2010, Weijer2019}. The reason is that this quantity is 
a measure of the salt-advection feedback strength, thought crucial in AMOC 
tipping. This feedback describes the amplification of a freshwater  perturbation 
in  the North  Atlantic through a weakening of the    AMOC  which  leads to less 
northward salt transport and hence amplification of the initial freshwater 
perturbation \cite{Marotzke2000,Peltier2014}. 

\begin{figure}[h!]

\hspace{-2cm}
\includegraphics{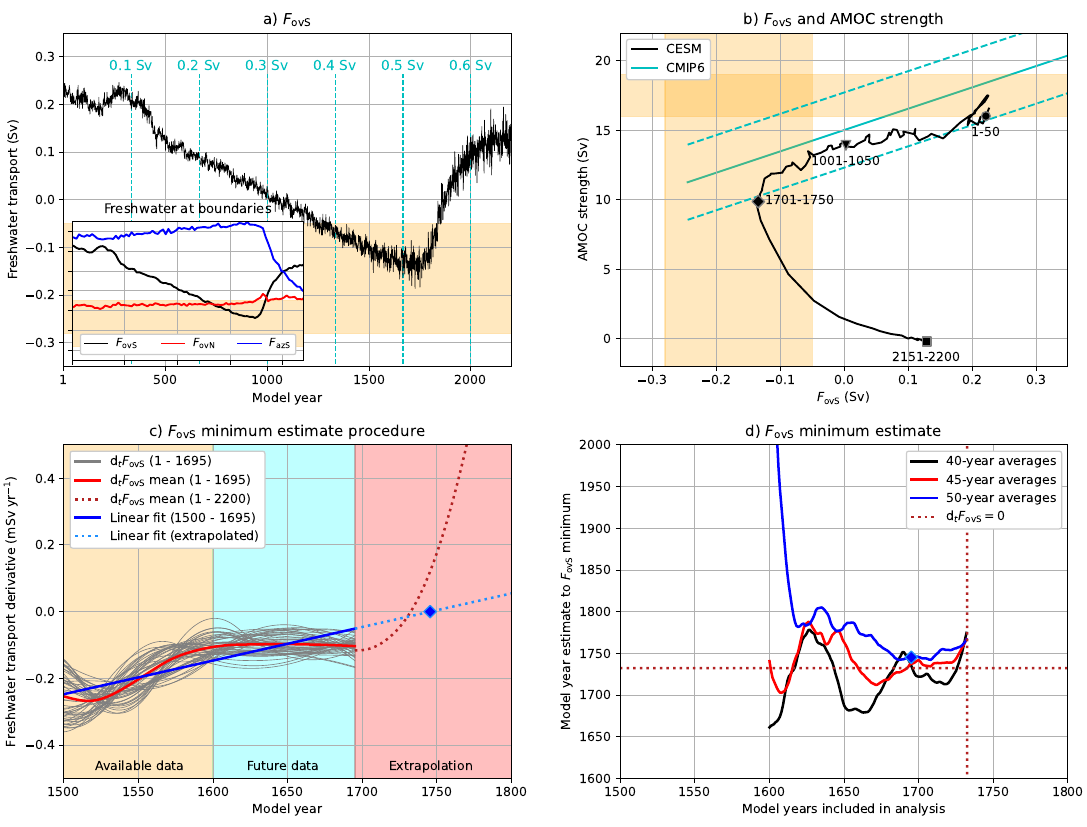}

\caption{\textbf{Freshwater transport by the AMOC and $F_{\mathrm{ovS}}$ minimum estimate.}
(a): The freshwater transport by the AMOC at 34$^{\circ}$S, $F_{\mathrm{ovS}}$.
The cyan-coloured lines indicate the magnitude of $F_H$.
Inset: Same dimensions as the main figure, but now for the freshwater transport at 34$^{\circ}$S ($F_{\mathrm{ovS}}$, black curve), 
60$^{\circ}$N ($F_{\mathrm{ovN}}$, blue curve) and the azonal (gyre) component at 34$^{\circ}$S ($F_{\mathrm{azS}}$, cyan curve).
The time series are displayed as 25-year averages (to reduce the variability of the time series).
(b): The $F_{\mathrm{ovS}}$ and AMOC strength over time, 
the time series are displayed as 25-year averages (to reduce the variability of the time series).
The markers indicate the 50-year average over a particular period.
The cyan-coloured curve indicates the present-day (1994 -- 2020) CMIP6 regression and 1 standard deviation \cite{vanWesten2023b}.
The yellow shading in panels~a and b indicates observed ranges \cite{Garzoli2013, Mecking2017, Smeed2018, Worthington2021} for $F_{\mathrm{ovS}}$ and AMOC strength.
(c): The $F_{\mathrm{ovS}}$ minimum estimate procedure (here for 50-year averages, see Methods). 
The time derivative of $F_{\mathrm{ovS}}$ ($\mathrm{d}_t F_{\mathrm{ovS}}$, red curve) is determined over the available and future data (up to model year~1,695), 
then a linear trend is determined (model years~1,500 -- 1,695) which is extrapolated to find the zero (diamond label).
(d): The $F_{\mathrm{ovS}}$ minimum estimate for varying model years (i.e., available and future data) and different averaging periods, the dotted lines and diamond label are similar to the ones from panel~c.}

\label{fig:Figure_4}
\end{figure}

In the CESM results here (Figure~\ref{fig:Figure_4}a), 
$F_{\mathrm{ovS}}$ is positive at the beginning of the simulation, which indicates 
that the AMOC exports net salinity (w.r.t. reference salinity of 35~g~kg$^{-1}$) out 
of the Atlantic.  This is not in agreement with observations \cite{Bryden2011, vanWesten2023b} 
which is a well-known bias in CMIP phase 3 \cite{Drijfhout2011}, phase 5 
\cite{Mecking2017} and phase 6  \cite{vanWesten2023b} models. In the CMIP phase~6 (CMIP6)
models, this bias is mainly due to large biases (compared to observations) in the 
freshwater flux over the Indian Ocean \cite{vanWesten2023b}. 

With increasing anomalous North Atlantic freshwater forcing, a greater salinity transport into (and/or 
greater fresh water export out of) the Atlantic Ocean is needed to balance the Atlantic's 
salinity budget  \cite{Rahmstorf1996} resulting in a declining $F_{\mathrm{ovS}}$.
Note that the AMOC becomes weaker under the freshwater forcing, so the salinity fields 
at 34$^{\circ}$S, which partly control $F_{\mathrm{ovS}}$, must change accordingly to 
balance the Atlantic's salinity budget (Figure~\ref{fig:Figure_S5}). The range of 
$F_{\mathrm{ovS}}$ and AMOC changes in the northward overturning regime 
(until model year~1,750) are within (Figure~\ref{fig:Figure_4}b) that of present-day 
simulations of CMIP6 models \cite{vanWesten2023b}. 
But the most important result here is that $F_{\mathrm{ovS}}$ goes 
through a minimum  very close to the collapse (near model year~1,750). 
Conceptual AMOC models \cite{Cessi1994, Rahmstorf1996} clearly identify such 
a minimum with a saddle-node bifurcation \cite{Stommel1961} which is the AMOC 
tipping point in these models. 

At the $F_{\mathrm{ovS}}$ minimum, an increment in the anomalous surface freshwater 
forcing weakens the AMOC further and this weakening can not induce sufficient salinity 
transport into (and/or fresh water transport out of) the Atlantic Ocean (hence $F_{\mathrm{ovS}}$ 
increases) resulting in  the $F_{\mathrm{ovS}}$ minimum. In other words, the dominant balance 
between  $F_H$ and $F_{\mathrm{ovS}}$ changes cannot be sustained under AMOC weakening.
The tipping point is at slightly higher values of $F_H$ (and hence slightly later in the 
simulation) than the value  at the $F_{\mathrm{ovS}}$ minimum, so the latter is a lower bound 
for tipping. This is  because the (azonal) gyre component at the southern 
boundary (34$^{\circ}$S, $F_{\mathrm{azS}}$) and the freshwater transport changes at the 
northern boundary (60$^{\circ}$N, $F_{\mathrm{ovN}}$) also contribute to the Atlantic Ocean's 
salinity budget \cite{Dijkstra2007}. Although these responses under the surface freshwater forcing 
are fairly small compared to $F_{\mathrm{ovS}}$ changes (inset in Figure~\ref{fig:Figure_4}a) 
they allow the existence of a near-equilibrium AMOC state for larger $F_H$ values than that 
at the   $F_{\mathrm{ovS}}$ minimum.  Eventually the ocean circulation and salinities adjust to 
the collapsed state (Figure~\ref{fig:Figure_S5}) and $F_{\mathrm{ovS}}$ becomes positive again, 
which is in line with the analyses from idealised climate model studies  \cite{Dijkstra2007, 
Huisman2010}. 

\newpage

The $F_{\mathrm{ovS}}$ minimum in our CESM simulation can be reasonably estimated almost 
100~years in advance (Figure~\ref{fig:Figure_4}d). We use cubic splines to  find a point in the `future' 
where  the time derivative of $F_{\mathrm{ovS}}$ (indicated by $\mathrm{d}_t F_{\mathrm{ovS}}$) 
goes  through zero, marking the $F_{\mathrm{ovS}}$ minimum (see Methods).  Using 100~years of data 
(i.e., Available data in Figure~\ref{fig:Figure_4}c) is insufficient for a reliable $F_{\mathrm{ovS}}$ 
minimum estimate, but including more data to the analysis (i.e, Future data in Figure~\ref{fig:Figure_4}c) 
improves the estimate in the CESM substantially. The result is also robust to the averaging interval, 
when longer than 35 years (Figure~\ref{fig:Figure_4}d). 

\begin{figure}[h]

\hspace{-2cm}
\includegraphics{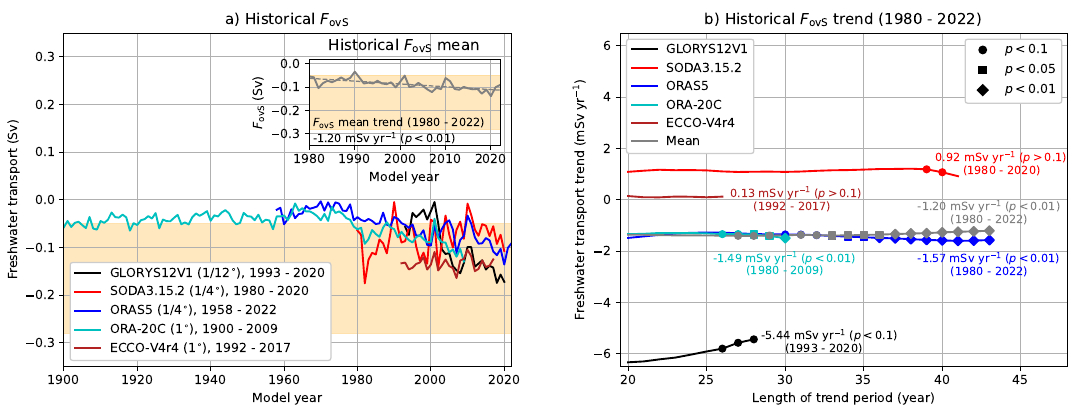}

\caption{\textbf{Historical $F_{\mathrm{ovS}}$ and trend.}
(a): The historical $F_{\mathrm{ovS}}$ for 5~different reanalysis and assimilation products. 
The horizontal resolution for the ocean component and the time span are indicated in the legend.
The $F_{\mathrm{ovS}}$ multi-reanalysis mean (inset) is the yearly average over all available products.
The yellow shading indicates observed ranges for  $F_{\mathrm{ovS}}$.
(b): The present-day (1980 -- 2022) $F_{\mathrm{ovS}}$ trend.
We use a sliding window (varying length of 20 to 43~years) over the available time series (1980 -- 2022) to determine all  $F_{\mathrm{ovS}}$ trends 
and then determine the mean trend, which is displayed for each reanalysis product and the multi-reanalysis mean.
The markers indicate the least significant (two-sided t-test \cite{Santer2000}) trend for a given sliding window and 
the trends at the maximum sliding window length (only one trend possible) are also displayed.}

\label{fig:Figure_5}
\end{figure}

The historical $F_{\mathrm{ovS}}$, derived from reanalysis and assimilation products 
(Figure~\ref{fig:Figure_5}a), are consistent in the sign of $F_{\mathrm{ovS}}$ when 
comparing those to direct observations \cite{Bryden2011, Garzoli2013}. 
The reanalysis product mean shows a robust and significant negative $F_{\mathrm{ovS}}$ trend (of -1.20~mSv~yr$^{-1}$) over the last 40~years (Figure~\ref{fig:Figure_5}b)
and its magnitude is close to the projected CMIP6 mean trend (of -1.06~mSv~yr$^{-1}$, 2000 -- 2100 \cite{vanWesten2023b}) under a high-end climate change scenario. This multi-reanalysis mean 
negative trend  suggests that the  AMOC is on course to  tipping. Although the reanalysis products
are known to have different biases \cite{Jackson2022}, this trend estimate  is the best result which 
can be obtained  at the moment.   However, these products are too short (maximum of 
$\sim$100~years) at the moment to adequately  estimate the distance to the  
$F_{\mathrm{ovS}}$  minimum. 

\begin{figure}[h]

\hspace{-2cm}
\includegraphics{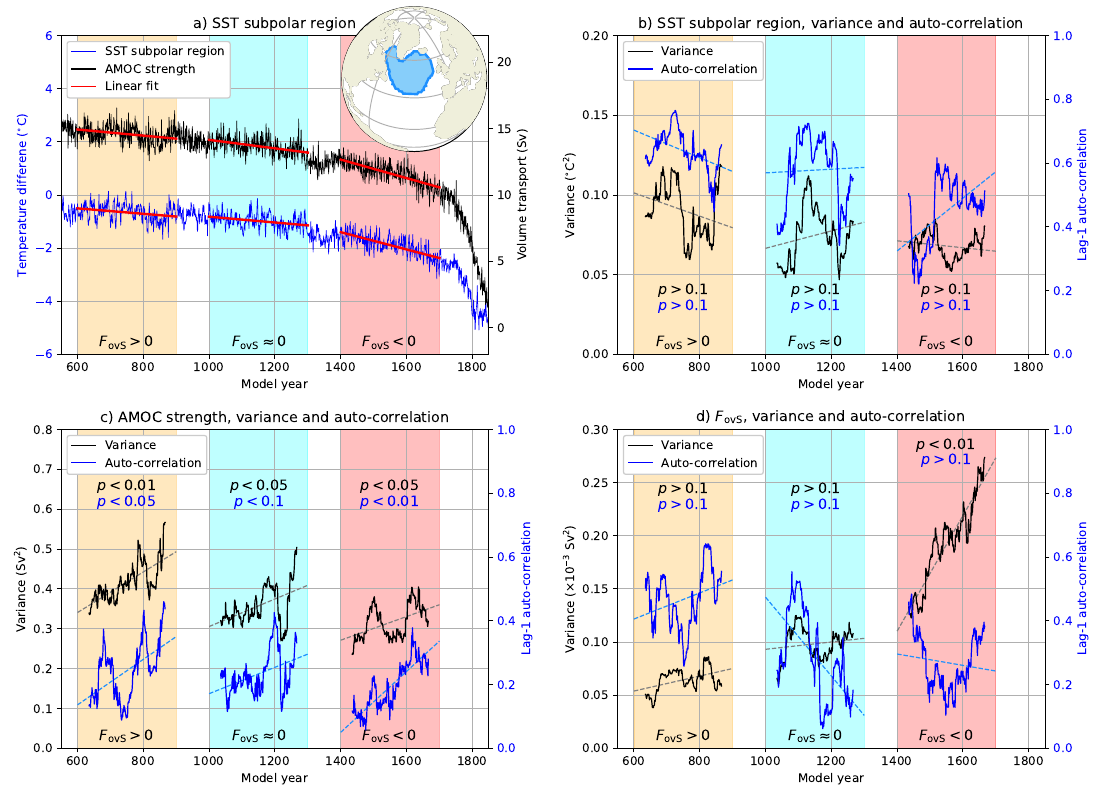}

\caption{\textbf{Classical early warning indicators.}
(a): The spatially-averaged SST over the subpolar region (blue outlined region in inset)
and AMOC strength (similar to Figure~\ref{fig:Figure_1}a) prior to the AMOC collapse.
The SST time series is averaged over the months November to May, where first the global mean SST (November to May  \cite{Ceasar2018}) and then the time mean over the first 50~years are subtracted.
(b): The variance and (lag-1) auto-correlation of the SST subpolar region for three 300-year periods. 
For each 300-year period, the linear trend is removed (red lines in panel~a) before determining the variance and auto-correlation.
The variance and auto-correlation are determined over a 70-year sliding window and the significance of their trends (dashed lines) are indicated for each period (see \cite{Boers2021} for more details).
(c \& d): Similar to panel~b, but now for the (c): AMOC strength (panel~a) and (d): the $F_{\mathrm{ovS}}$ (Figure~\ref{fig:Figure_4}a).}

\label{fig:Figure_6}
\end{figure}

By  analysing SST-based proxies of the AMOC strength \cite{Ceasar2018}, it was suggested that the real 
present-day AMOC approaches a tipping point \cite{Boers2021, Ditlevsen2023}. 
Following the same procedure as outlined in \cite{Boers2021} on our AMOC tipping event in the CESM, 
we find no consistent increase in the  classical early warning indicators (i.e., variance and lag-1 auto-correlation) 
prior to the AMOC  collapse (Figures~\ref{fig:Figure_6}a,b). The early warning indicators (temporarily and 
non-significantly) can increase  within each 300-year period indicated (shaded intervals), but over longer 
time scales these quantities decline when approaching  the tipping point. 
Conducting the same analysis over the last 150~years (model years~1,550 -- 1,700, 50~years before AMOC 
tipping event) indicates that the inter-annual variability, variance ($\approx$ 0.05 -- 0.07$^{\circ}$C$^2$) and 
auto-correlation ($\approx$ 0.4 -- 0.6) are very similar between the CESM and the observed SST time series 
(Figure~4 in \cite{Boers2021}). The variance is also increasing ($p < 0.1$) over this 150-year period, but the 
auto-correlation is decreasing ($p > 0.1$). 
For model years~1,427 -- 1,557 an accurate tipping point estimate (following the procedure outlined in 
\cite{Ditlevsen2023}) can be made which is consistent with the timing of the  $F_{\mathrm{ovS}}$ minimum (red 
curves in Figure~\ref{fig:Figure_S6}). This estimate is only accurate  when both the variance and auto-correlation 
increase. For various 150-year periods (and when  $F_{\mathrm{ovS}} < 0$)  either variance or auto-correlation  
does not increase,  resulting  in inaccurate  (blue curves in Figure~\ref{fig:Figure_S6}) tipping point estimates  (Figure~\ref{fig:Figure_S6}d). 
Because both  variance and autocorrelation are increasing  in the SST-based 
time series in  \cite{Ditlevsen2023}, their estimate of the tipping point (2025 -- 2095, 95\%-confidence level) could be 
accurate. On the other hand, our results (Figure~\ref{fig:Figure_S6}) also show that the accuracy is 
sensitive to the time interval analysed, due to decadal variability in the SST time series,  and that most 
150-year time intervals do not provide an accurate estimate of the tipping point. 

Using the classical early warning indicators directly on the AMOC 
strength time series (Figure~\ref{fig:Figure_6}c) gives similar results as the SST time series.
The trends  over the 300-year periods are significant but the overall variance and auto-correlation are decreasing towards the tipping point.
Only for the $F_{\mathrm{ovS}}$ (Figure~\ref{fig:Figure_6}d), we find a consistent and significant increase in 
its variance when approaching the tipping point and this can also be directly observed from 
the full time series (Figure~\ref{fig:Figure_4}a). The quantity $F_{\mathrm{ovS}}$, in particular its
minimum in  combination with its variance increase, is hence a very promising  early warning 
signal for a (future) AMOC collapse.

\section*{Discussion}

The results in this paper give a clear answer to a long-standing problem 
around in the  climate research community concerning the existence 
of AMOC tipping behaviour in state-of-the-art GCMs 
\cite{Stouffer2006, Mecking2016, Jackson2018a, Liu2014, Jackson2018b, 
Weijer2019}.  Yes, it does occur in these models!  This  is bad news for the 
climate system and humanity  as up till now one could think that AMOC tipping 
was  only a theoretical  concept and tipping would disappear as soon as the full 
climate  system, with all its additional feedbacks, was considered. On the 
other hand, the tipping is consistent with the wealth on paleo-climate 
evidence that rapid changes have occurred in the AMOC, in particular 
during Dansgaard-Oeschger events  \cite{Lynch-Stieglitz2017}.

The AMOC collapse dramatically changes the redistribution of heat (and salt) 
and results in a cooling of the Northern Hemisphere while the Southern Hemisphere 
slightly warms. Atmospheric and sea-ice feedbacks, which were not considered in 
idealised climate models studies  \cite{Cessi1994, Rahmstorf1996, Dijkstra2007, 
Huisman2010}, further amplify the AMOC-induced changes resulting in a very strong 
and rapid cooling of the European climate with temperatures trends of more than 
3$^{\circ}$C per decade. In comparison with the present-day global mean surface temperature 
trend (due to climate change) of about 0.2$^{\circ}$C per decade, no realistic adaptation 
measures can deal with such rapid temperature changes under an AMOC collapse  
\cite{Vogel2004, Birkmann2010}.

We have developed a physics-based, and observable 
\cite{Bryden2011, Garzoli2013},  early warning signal characterising
the tipping point of the AMOC: the minimum of the AMOC induced 
freshwater transport at 34$^\circ$S in the Atlantic, here indicated by 
$F_{\mathrm{ovS}}$. 
The quantity $F_{\mathrm{ovS}}$ has  a strong basis in conceptual models,  where it is  an 
indicator of the salt-advection feedback strength.   Although 
$F_{\mathrm{ovS}}$ has been shown to be a useful measure of AMOC stability 
in GCMs \cite{Jackson2013}, the minimum feature has so far not been connected 
to the tipping point because an AMOC tipping event had up to now not been found in
these models.  The $F_{\mathrm{ovS}}$ indicator is observable and reanalysis 
products show that its value and, more importantly its trend,  are negative  at the 
moment. This provides a strong  indication that the AMOC is on route towards 
tipping. 

In addition,  with future observations,   an estimate of the distance  to the AMOC tipping point
can in principle be obtained. Deploying machine learning techniques on $F_{\mathrm{ovS}}$, 
in combination with its variance, could also help in estimating the distance to AMOC tipping.
We have shown that   current reanalysis products provide insufficient information 
to adequately estimate this  distance.  Sustained future section measurements (available since 
2009) at 34$^{\circ}$S  from the SAMBA  array \cite{Meinen2018, Kersale2020, 
Kersale2021} are  therefore of utmost  importance and will become crucial to estimate 
the distance to an AMOC collapse.  

In the CESM simulation here, AMOC tipping occurs at relatively large values
of the  freshwater forcing. This is due to  biases in precipitation 
elsewhere  in the models and mainly over the Indian Ocean 
\cite{vanWesten2023b}. Hence, we needed to integrate the CESM to 
rather  large  values of the freshwater forcing ($\sim0.6$~Sv, about a 
factor 80~times larger than the present-day melt rate of the Greenland Ice 
Sheet \cite{Sasgen2020}) to find  the AMOC tipping event. The  effect of the 
biases can be seen from the value  of the AMOC induced freshwater 
transport at 34$^\circ$S, $F_{\mathrm{ovS}}$,  which is positive at the 
start of the simulation.    When biases are corrected in the CESM, it is  expected that the 
AMOC tipping  is expected to occur at smaller values of the  freshwater 
forcing. As also the present-day background climate state and the climate 
change forcing are different than in our simulations,  the real present-day 
AMOC may be much  closer to its tipping point than in the simulations 
shown here. Note that the analysis of the early warning signal
is not affected by these biases as this analysis is independent of the 
background state and precise forcing details.

\section*{Methods}

\textbf{Climate Model Simulations} 
The hosing experiment was branched off from the end (model year~2,800) of the 
pre-industrial CESM control simulation from Baatsen et al. \cite{Baatsen2020}. Here
it is  shown that  the upper 1,000~m of the ocean is well equilibrated after 2,800~years 
of model integration. The horizontal resolutions are 1$^{\circ}$ for the ocean/sea ice  
and 2$^{\circ}$ for the atmosphere land model components in CESM. 
We added a freshwater flux anomaly over the North Atlantic ocean surface hosing region 
between  20$^{\circ}$N and  50$^{\circ}$N (inset in Figure~\ref{fig:Figure_1}a). To conserve 
salinity, this freshwater flux anomaly was compensated over  the ocean surface outside 
the hosing region.  We linearly increased the freshwater forcing with a trend $3 \times 
10^{-4}$~Sv~yr$^{-1}$ to a maximum of $F_H = 0.66$~Sv in model year~2,200.  \\ 

\textbf{The Freshwater Transport} 
The freshwater transport by the overturning component ($F_{\mathrm{ovS}}$) and azonal (gyre) component ($F_{\mathrm{azS}}$) at 34$^{\circ}$S are determined as: 

\begin{subequations}
\begin{align}
F_\mathrm{ovS} = F_\mathrm{ov}(y = 34^{\circ}\mathrm{S}) &= - \frac{1}{S_0} \int_{-H}^0 \left[ \int_{x_W}^{x_E} v^* \mathrm{d} x \right] \left[ \langle S \rangle - S_0 \right] \mathrm{d}z \\
F_\mathrm{azS} = F_{\mathrm{az}}(y = 34^{\circ}\mathrm{S})  &= - \frac{1}{S_0} \int_{-H}^0 \int_{x_W}^{x_E} v' S' \mathrm{d} x \mathrm{d}z 
\end{align}
\end{subequations}

where $S_0 = 35$~g~kg$^{-1}$ is a reference salinity. The $v^*$ indicates the baroclinic velocity and is defined as $v^* = v - \hat{v}$,
where $v$ is the meridional velocity and $\hat{v}$ the barotropic meridional velocity (i.e., the section spatially-averaged meridional velocity).
The $\langle S \rangle$ indicate the zonally-averaged salinity and primed quantities ($v'$ and $S'$) are deviations from their respective zonal means. 
For more details we refer to J\"uling et al. (2021) \cite{Juling2021}.\\

\textbf{The AMOC strength} 
The AMOC strength is defined as the total meridional volume transport at 26$^{\circ}$N over the upper 1,000~m:
 \begin{equation} \label{eq:AMOC}
\mathrm{AMOC}(y = 26^{\circ}\mathrm{N}) = \int_{-1000}^{0} \int_{x_W}^{x_E} v~\mathrm{d}x \mathrm{d}z
 \end{equation}
\\

\textbf{$F_{\mathrm{ovS}}$ minimum estimate} To estimate the $F_{\mathrm{ovS}}$ minimum,
we use cubic splines that interpolate piecewise, between so-called knots, cubic polynomials which are twice continuously differentiable and we impose that the second derivate is zero at the first and last knot. 
The knots are determined over $n$-year averages of the $F_{\mathrm{ovS}}$ time series for different starting years (1, 2, ..., $n-1$) 
and result in $n$ different cubic splines and their respective derivatives ($\mathrm{d}_t F_{\mathrm{ovS}}$, grey curves in Figure~\ref{fig:Figure_4}c).
Using a linear fit over the cubic spline mean derivative, we estimate where the derivative goes through zero (i.e., the $F_{\mathrm{ovS}}$ minimum).
A minimum of $n \geq 35$ (year averages and cubic splines) is required to substantially reduce the variability of the time series and find a consistent $F_{\mathrm{ovS}}$ minimum estimate. \\

\section*{Software and Model Output} 
The (processed) model output and analysis scripts will be made available on Zenodo upon publication.
The reanalysis and assimilation products can be accessed through: GLORYS12V1 (https://doi.org/10.48670/moi-00021), SODA3.15.2 (http://www.soda.umd.edu)
ORAS5 (https://doi.org/10.24381/cds.67e8eeb7), \\
ORA-20C (https://icdc.cen.uni-hamburg.de/thredds/catalog/ftpthredds/EASYInit/ora20c/opa0/catalog.html)
and ECCO-V4r4 (https://www.ecco-group.org/products-ECCO-V4r4.htm). \\

\section*{Acknowledgements}

The model simulation and the analysis of all the model output was conducted on the Dutch National 
Supercomputer Snellius within NWO-SURF project 17239. 
R.M.v.W. and H.A.D. are funded by the European Research Council through the ERC-AdG project TAOC (project 101055096). 

\section*{Author Contributions Statement}

R.M.v.W. and H.A.D. conceived the idea for this study. M.K. performed the model simulation with the CESM.
R.M.v.W. conducted the analysis and prepared all figures. All authors were actively involved in the interpretation of the analysis results and the writing process.






\bibliographystyle{Science}


\setcounter{figure}{0}

\makeatletter 
\renewcommand{\thefigure}{S\@arabic\c@figure}
\makeatother

\begin{figure}[h]

\hspace{-2cm}
\includegraphics{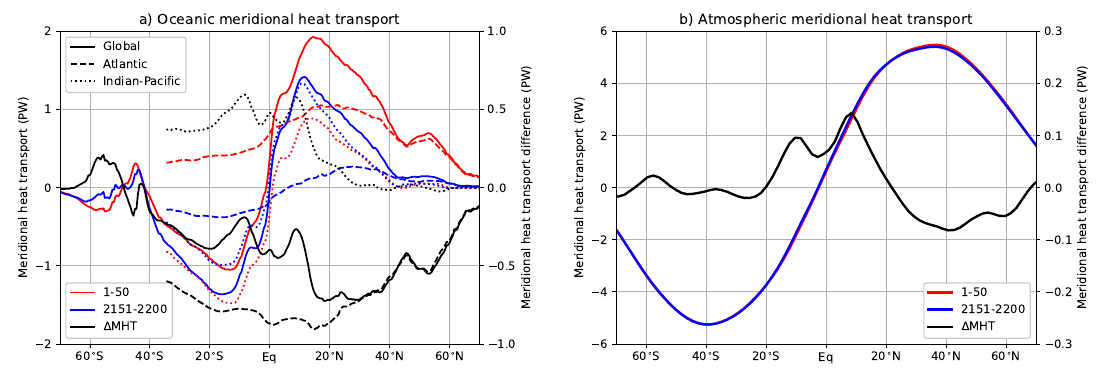}

\caption{\textbf{Meridional heat transport}
(a \& b): The oceanic and atmospheric meridional heat transport (MHT) for model years~1 -- 50 (red curves) and 2,151 -- 2,200 (blue curves) and differences (black curves) between the two periods.
In panel~a the meridional heat transport is shown for the global ocean (solid curves), Atlantic Ocean (dashed curves) and Indian-Pacific Ocean (dotted curves).
Note the different vertical ranges between the two panels.}

\label{fig:Figure_S1}
\end{figure}


\begin{figure}[h]
\hspace{-2cm}
\includegraphics{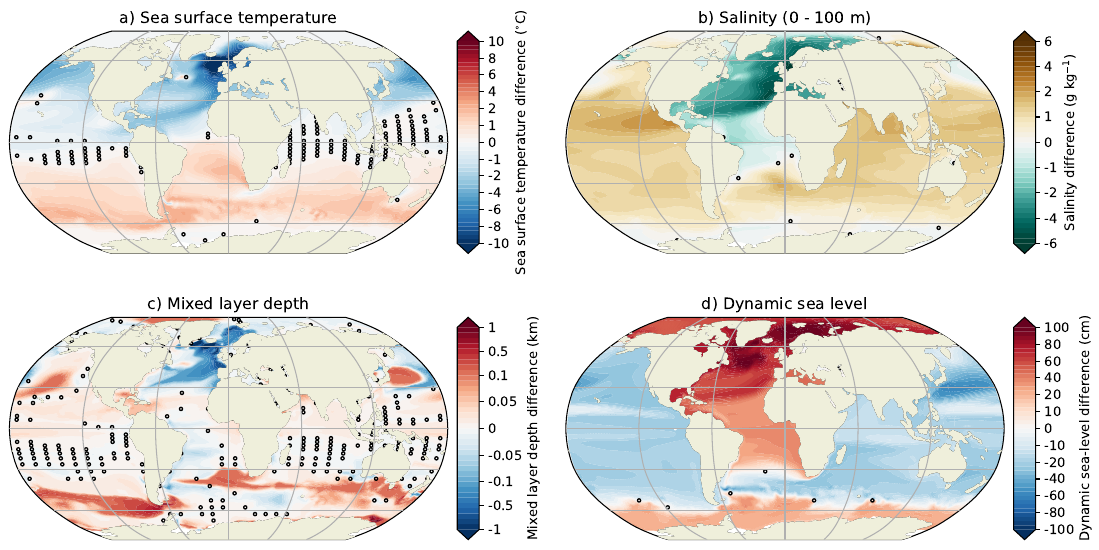}
\caption{\textbf{Oceanic response.}
(a): The sea surface temperature differences between the two AMOC states (model years~2,151 -- 
2,200 minus 1 -- 50), the markers indicate non-significant ($p \geq 0.05$, two-sided Welch's t-test) 
differences. (b -- d): Similar to panel~a, but now for the (b): vertically-averaged 
(0 -- 100~m) salinity, (c): yearly-maximum mixed layer depth and (d): dynamic sea level.}
\label{fig:Figure_S2}
\end{figure}


\begin{figure}[h]

\vspace{-2cm}
\hspace{-2cm}
\includegraphics{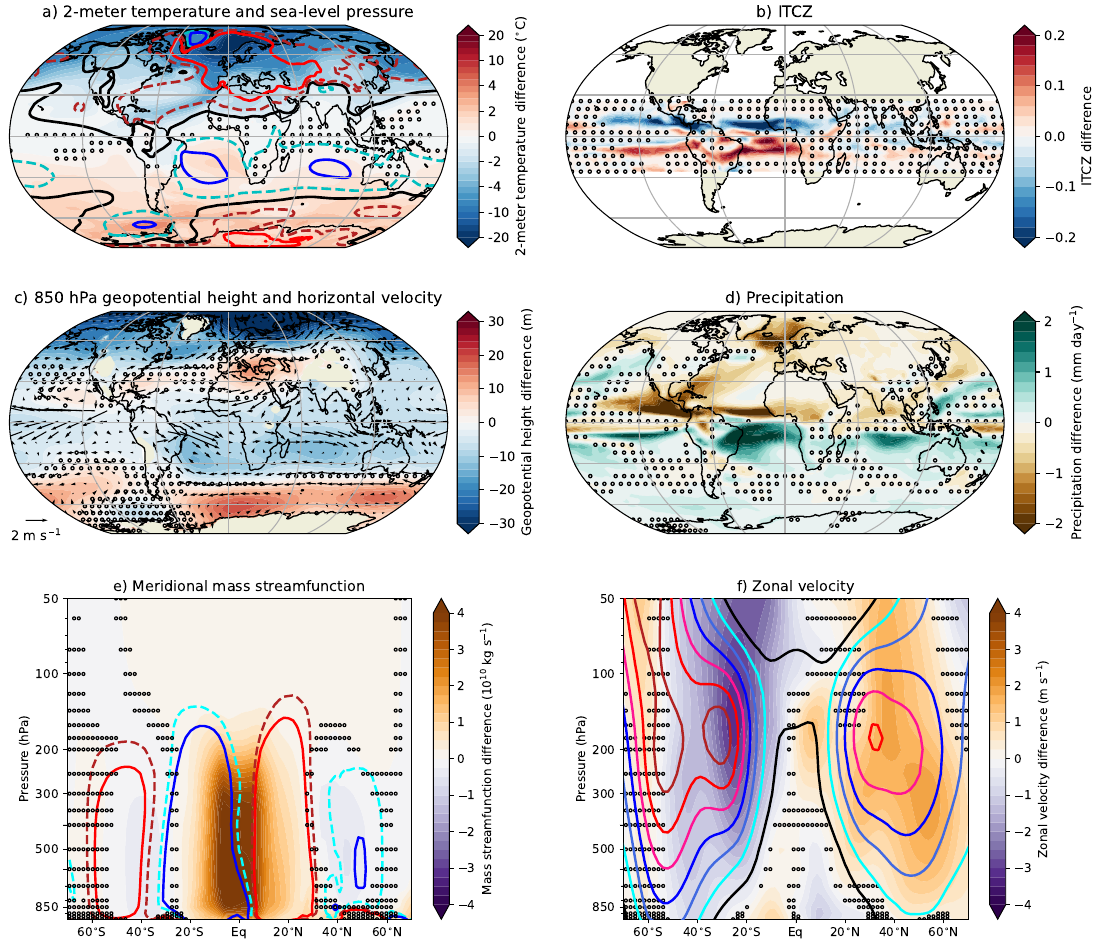}

\caption{\textbf{Atmospheric response.}
(a): The 2-meter surface temperature differences between the two AMOC states (model years~2,151 -- 2,200 minus 1 -- 50), 
the markers indicate non-significant ($p \geq 0.05$, two-sided Welch's t-test) differences. 
The red (blue) curves show positive (negative) values of sea-level pressure differences with magnitudes of (-)1~hPa
and (-)2~hPa for the dashed and solid curves, respectively.
(b -- f): Similar to panel~a, but now for the (b): ITCZ location, (c): 850~hPa geopotential height (shading) and 850~hPa horizontal velocities (quivers), (d): precipitation.
(e): meridional mass streamfunction and (f): zonally-averaged zonal velocity.
The ITCZ location is determined from the monthly-averaged joint distribution of outgoing longwave radiation and precipitation and are then converted to yearly probabilities \cite{Mamalakis2021}.
The curves in panel~e show the meridional mass streamfunction for model years~1 -- 50,
where the red (blue) curves are positive (negative) value with magnitudes of
$(-)1\times10^{10}$~kg~s$^{-1}$ and  $(-)2\times10^{10}$~kg~s$^{-1}$ for the dashed and solid curves, respectively.
The curves in panel~f show the zonal velocity for model years~1 -- 50, where the black curve is 0~m~s$^{-1}$ and the coloured curves are spaced every +5~m~s$^{-1}$.
}

\label{fig:Figure_S3}
\end{figure}


\begin{figure}[h]
\hspace{-2cm}
\includegraphics{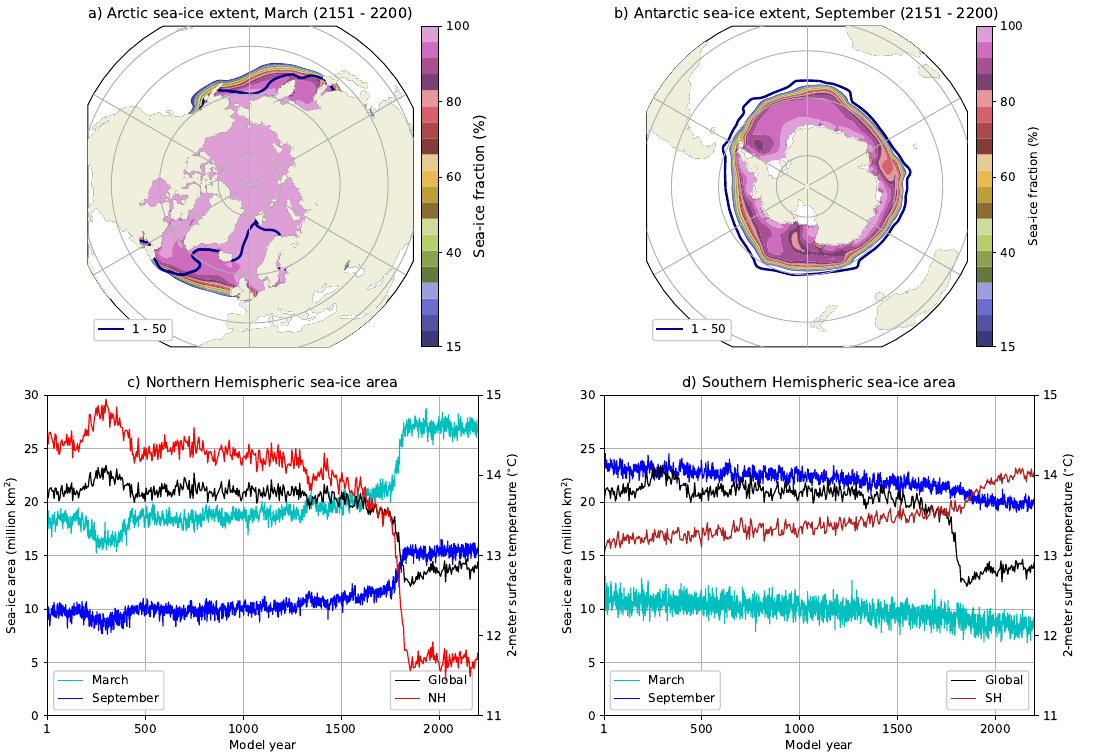}
\caption{\textbf{Sea-ice response.}
(a): The Arctic sea-ice fraction for March for model years~2,151 -- 2,200.
The dark blue curve shows the Arctic sea-ice edge (i.e, the 15\% sea-ice fraction isoline)  for March for model years~1 -- 50.
(b): Similar to panel~a, but now for the Antarctic sea-ice fractions for September.
(c): The Northern Hemispheric sea-ice area for March and September, including the 2-meter surface temperature for the global mean and Northern Hemisphere.
The sea-ice area is based on all the grid cells with sea-ice fractions larger than 15\%.
The 2-meter surface temperature time series are displayed as 5-year averages (to reduce the variability of the time series).
(d): Similar to panel~c, but now for the Southern Hemispheric sea-ice area and the Southern Hemispheric 2-meter surface temperature.
}

\label{fig:Figure_S4}
\end{figure}


\begin{figure}[h]

\hspace{-2.5cm}
\includegraphics[width=1.3\columnwidth]{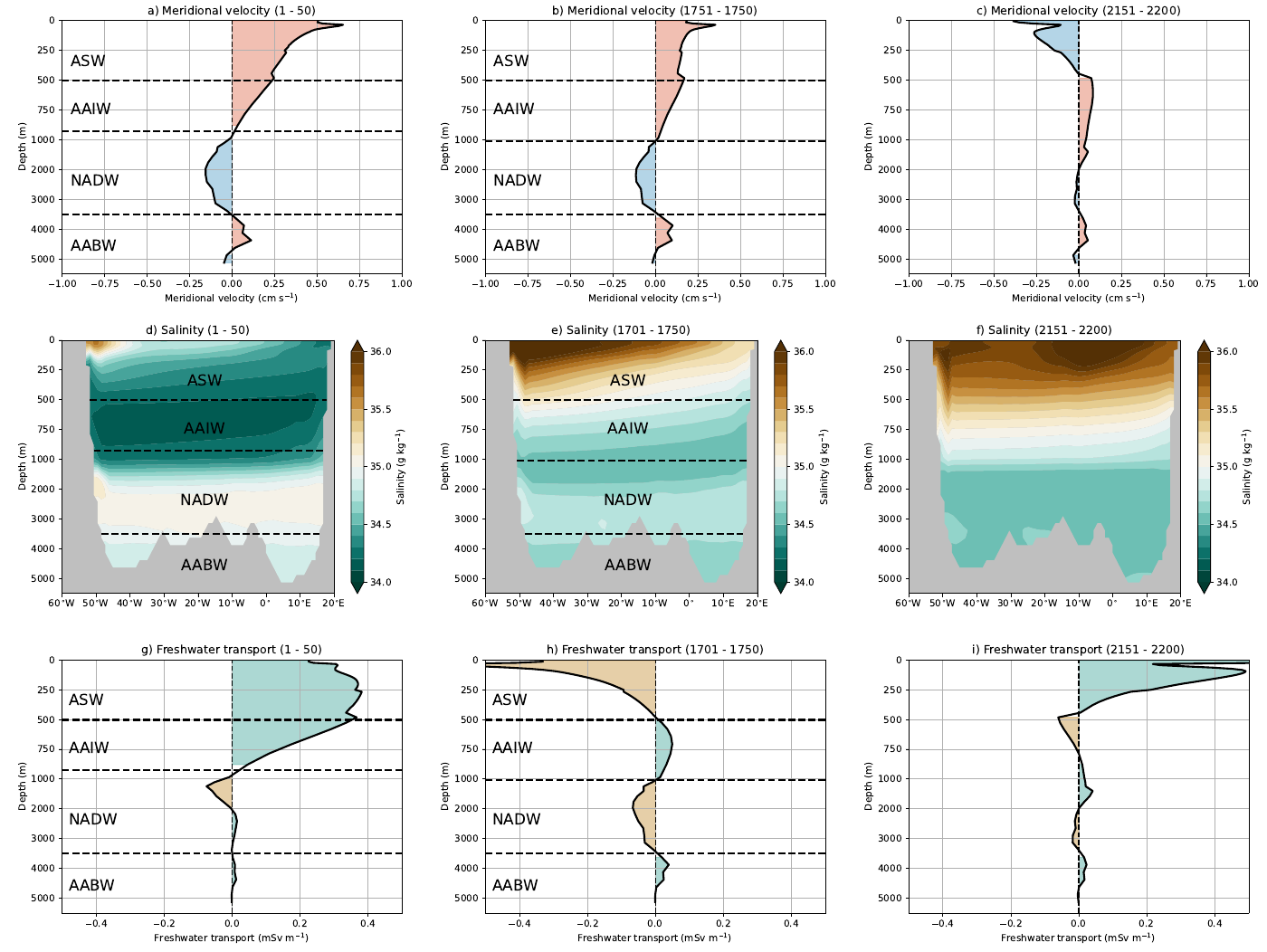}

\caption{\textbf{Water mass properties at 34$^{\circ}$S.}
(Upper row): The zonally-averaged meridional velocity at 34$^{\circ}$S for three periods (model years 1 -- 50, 1,701 -- 1,750 and 2,151 -- 2,200.).
(Middle row): The salinity along 34$^{\circ}$S  for the three periods.
(Lower row): The freshwater transport with depth at 34$^{\circ}$S for the three periods.
The different water masses are derived from the velocity profile \cite{vanWesten2023b} and is only applicable for the northward overturning circulation (left and middle column) and
the names are: Atlantic Surface Water (ASW), Antarctic Intermediate Water (AAIW), North Atlantic Deep Water (NADW) and Antarctic Bottom Water (AABW).}

\label{fig:Figure_S5}
\end{figure}


\begin{figure}[h]

\hspace{-2.5cm}
\includegraphics[width=1.3\columnwidth]{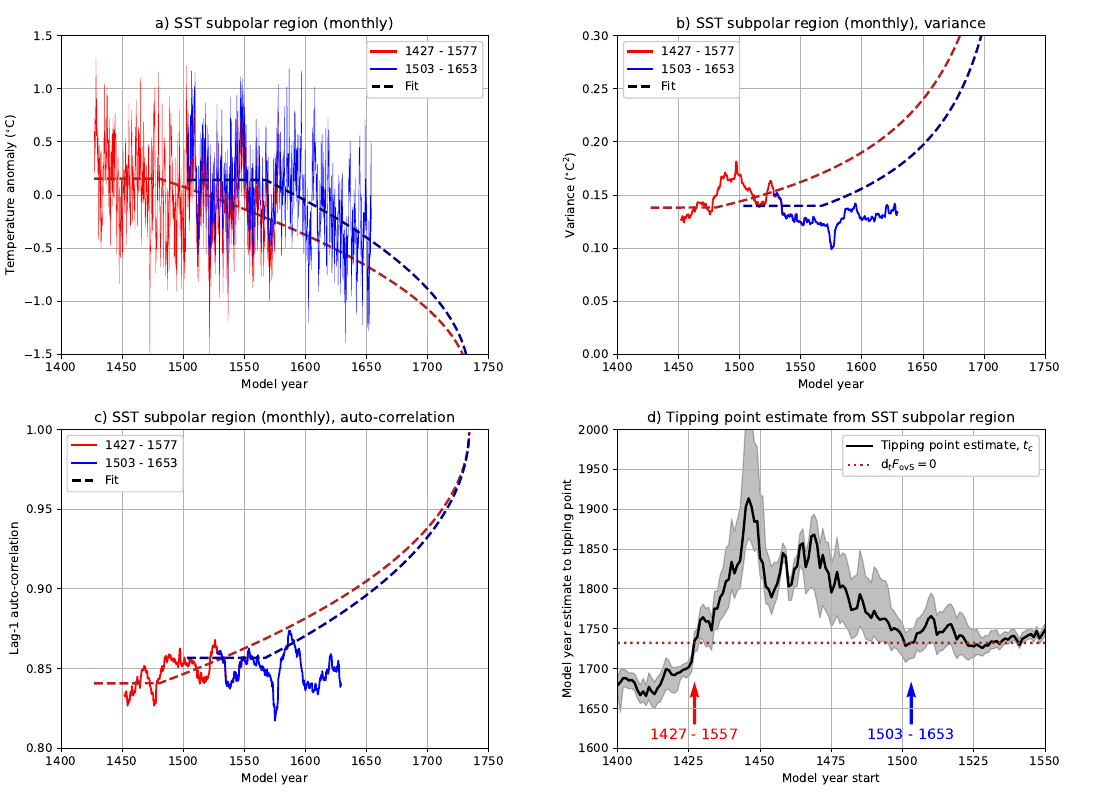}

\caption{\textbf{Tipping point estimate from SST subpolar time series.}
(a): The monthly SST subpolar time series (solid curves) for two 150-year periods. 
For each 150-year period, the monthly SST over the subpolar region (blue outlined region in inset in Figure~\ref{fig:Figure_6}a) and the monthly globally-averaged SST are determined.
Next, the monthly mean (i.e., January months, February months, etc.) is removed for both time series and then the globally-averaged SST anomaly time series 
is subtracted from the SST subpolar region anomaly time series.
We did not subtract twice the globally-averaged SST \cite{Ditlevsen2023} as there is no climate change scenario in the CESM.
(b \& c): The variance and lag-1 auto-correlation (solid curves) of the monthly SST subpolar time series over a 50-year sliding window.
A linear trend is removed over each sliding window before determining the variance and auto-correlation.
(d): The tipping point estimate, $t_c$, from the procedure outlined in \cite{Ditlevsen2023} over 150-year periods and varying starting year.
An example of the tipping point estimate (auto-correlation approaches~1) is shown by the dashed curves in panels~a -- c.
The shading indicates the 95\%-confidence interval and is determined by varying the length of the stationary part (50 to 80~years).}

\label{fig:Figure_S6}
\end{figure}


\end{document}